\documentclass{article}

\usepackage{booktabs}
\usepackage{graphics}
\usepackage{graphicx}
\usepackage{listings}
\usepackage{nameref}
\usepackage{algorithm}
\usepackage{algpseudocode}
\algrenewcommand\algorithmicrequire{\textbf{Input:}}
\algrenewcommand\algorithmicensure{\textbf{Output:}}
\usepackage{bm}
\usepackage{amsmath}
\usepackage{amssymb}
\usepackage{caption} 
\usepackage{multirow}
\usepackage{color}
\usepackage[table]{xcolor}

\usepackage{rotating}

\usepackage{arxiv}

\usepackage[utf8]{inputenc} 
\usepackage[T1]{fontenc}    
\usepackage{hyperref}       
\usepackage{url}            
\usepackage{booktabs}       
\usepackage{amsfonts}       
\usepackage{nicefrac}       
\usepackage{microtype}      
\usepackage{lipsum}
\usepackage{graphicx}
\graphicspath{ {./images/} }

\begin{document}
\title{Computational approaches for virus host prediction: A review of methods and applications}

\author{
 Jiayu Shang \\
  Dept. of Information Engineering\\
  Chinese University of Hong Kong\\
  Kowloon, Hong Kong SAR, China\\
  \And
 Cheng Peng \\
  Dept. of Electrical Engineering\\
  City University of Hong Kong\\
  Kowloon, Hong Kong SAR, China\\
  \And
 Jiaojiao Guan \\
  Dept. of Electrical Engineering\\
  City University of Hong Kong\\
  Kowloon, Hong Kong SAR, China\\
  \And
 Dehan Cai \\
  Dept. of Electrical Engineering\\
  City University of Hong Kong\\
  Kowloon, Hong Kong SAR, China\\
  \And
 Donglin Wang \\
  Sch. of Environmental Science and Engineering\\
  Shandong University\\
  Qingdao, Shandong, China\\
  \And
 Yanni Sun \\
  Dept. of Electrical Engineering\\
  City University of Hong Kong\\
  Kowloon, Hong Kong SAR, China\\
}

\maketitle

\begin{abstract}
Accurate prediction of virus-host interactions is critical for understanding viral ecology and developing applications like phage therapy. However, the growing number of computational tools has created a complex landscape, making direct performance comparison challenging due to inconsistent benchmarks and varying usability. Here, we provide a systematic review and a rigorous benchmark of 27 virus-host prediction tools. We formulate the host prediction task into two primary frameworks—link prediction and multi-class classification—and construct two benchmark datasets to evaluate tool performance in distinct scenarios: a database-centric dataset (RefSeq-VHDB) and a metagenomic discovery dataset (MetaHiC-VHDB). Our results reveal that no single tool is universally optimal. Performance is highly context-dependent, with tools like CHERRY and iPHoP demonstrating robust, broad applicability, while others, such as RaFAH and PHIST, excel in specific contexts. We further identify a critical trade-off between predictive accuracy, prediction rate, and computational cost. This work serves as a practical guide for researchers and establishes a standardized benchmark to drive future innovation in deciphering complex virus-host interactions.
\end{abstract}

\section{Introduction}
\label{sec:intro}
Viruses are obligate intracellular parasites that require a host cell to replicate. A significant subset of viruses, known as bacteriophages or phages, exclusively infect and replicate within prokaryotic hosts, including both bacteria and archaea \cite{naureen2020bacteriophages}. Phages represent the vast majority of all known viruses and are the most abundant biological entities on Earth \cite{batinovic2019bacteriophages}, with an estimated population exceeding $10^{31}$ particles—a number greater than all other organisms combined \cite{cobian2016viruses}. This extraordinary abundance is matched by a remarkable diversity in their virion morphologies and genomic structure \cite{dion2020phage}. While tailed phages with double-stranded DNA (dsDNA) genomes are the most frequently studied, phages can also possess double-stranded RNA (dsRNA), single-stranded DNA (ssDNA), or single-stranded RNA (ssRNA) genomes \cite{zrelovs2020motley}. Their genome sizes are also remarkably varied, ranging from a few kilobase pairs (kbp) to those of megaphages, which can exceed 500 kbp \cite{al2020clades}. Phages are found in all ecosystems colonized by prokaryotes, including aquatic environments \cite{chen2020large}, terrestrial soils \cite{jansson2023soil}, and the human microbiome \cite{wei2024bacteriophages}. Within these microbial communities, phages function as key ecological and evolutionary drivers by shaping community structure, promoting host co-evolution, and mediating horizontal gene transfer \cite{naureen2020bacteriophages, huang2024adaptive}.

\begin{figure*}
    \centering
    \vspace{-0.3cm}
    \includegraphics[width=0.65\linewidth]{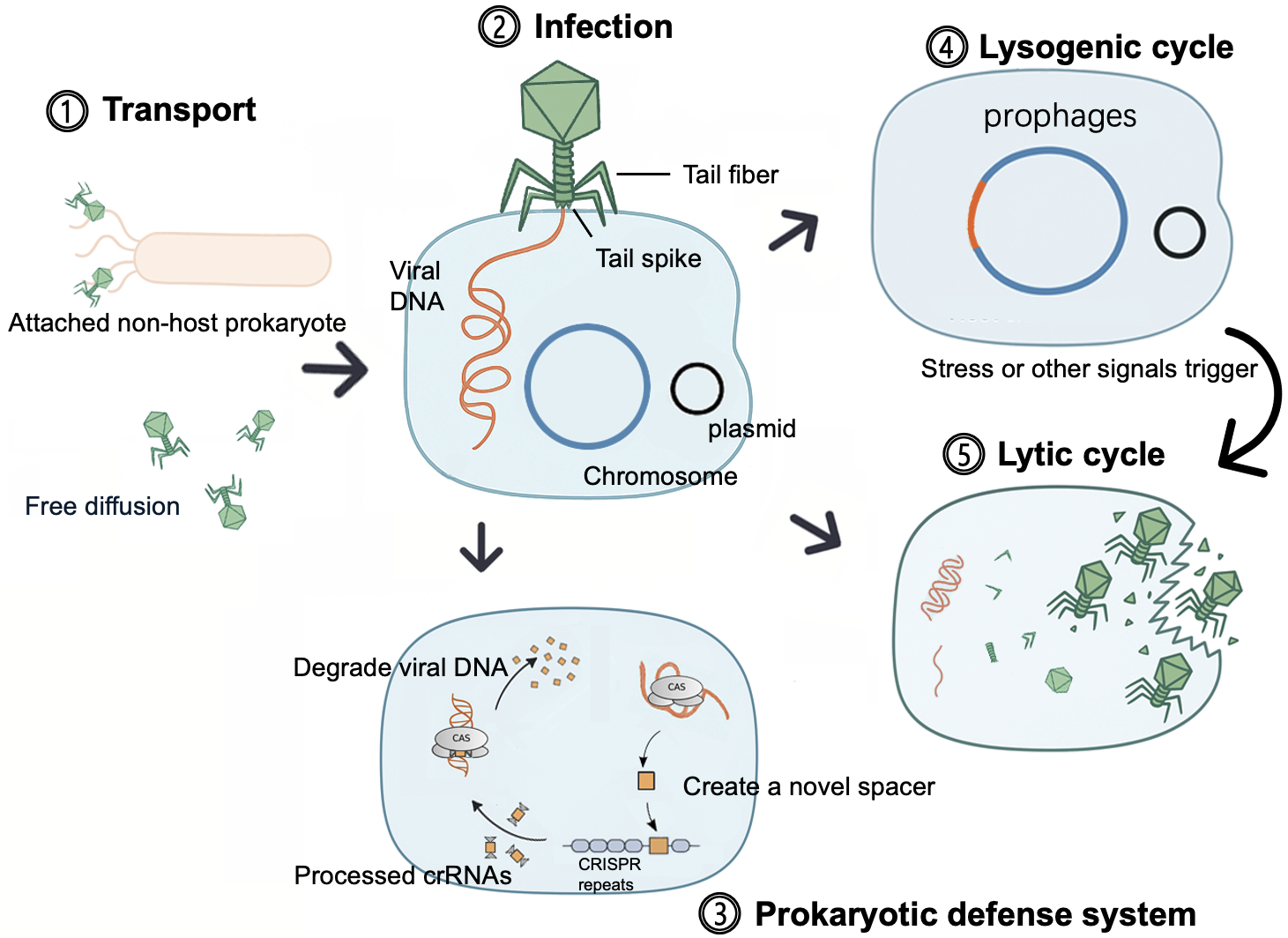}
    \caption{A schematic representation of the viral infection. (1) transport, where the virus attaches to a non-host prokaryote or diffuses freely. (2) infection, where the viral tail fiber and tail spike attach to the host and viral DNA is injected into the host cell. (3) The host can respond through its prokaryotic defense system, including CRISPR-Cas mechanisms that create novel spacers for immunity and degrade viral DNA. If the virus infect successfully, the virus may enter the (4) lysogenic cycle, integrating as prophages into the host chromosome, or (5) the lytic cycle, producing viral particles and lysing the host cell.}
    \vspace{-0.2cm}
    \label{fig:intro}
\end{figure*}

The relationship between a phage and a bacterium is not a simple predator-prey dynamic but rather a spectrum of possible interactions with distinct ecological consequences. A schematic representation of the viral infection is shown in Fig. \ref{fig:intro}. The process initiates with adsorption, where a phage recognizes and binds to specific receptors on a bacterial cell \cite{letarov2017adsorption, leprince2023phage}. This step is a critical checkpoint, but interactions are not limited to viable hosts; recent work shows that phages can leverage the flagellum of non-host bacteria as a means of transport \cite{yu2020hitchhiking, you2022mycelia}, enhancing their dispersal. Following successful injection of its genome, the phage must overcome the host's defense systems. Bacteria employ numerous anti-phage systems, including the well-studied CRISPR-Cas, restriction-modification, and methylation-associated defense systems, which provide an adaptive immune memory to fight off repeat infections \cite{hobbs2024nucleotide}. If the phage survives, the outcome is dictated by its lifestyle and reproductive strategy. Virulent phages engage in a lytic cycle, which involves hijacking host resources for viral replication and concludes with the destruction of the host cell to release progeny \cite{brady2021molecular}. In contrast, temperate phages may adopt a lysogenic lifestyle, integrating their DNA into the host genome as a prophage. In this state, the phage remains latent and is passively propagated with the host cell, a relationship that can persist until stress or other signals trigger its excision and entry into the lytic cycle \cite{howard2017lysogeny}.

The destructive potential of the lytic cycle, a key outcome of phage-bacteria interactions, provides the foundation for several critical biotechnological applications. The most prominent of these is phage therapy \cite{lin2017phage}, which is re-emerging as a compelling alternative to conventional antibiotics, especially in the face of rising antimicrobial resistance. Because phages are highly specific to their bacterial targets, some lytic phages can eliminate pathogens with minimal disruption to the patient's beneficial microbiota, a significant advantage over broad-spectrum antibiotics. This precision has shown promise in treating challenging infections, including those caused by multidrug-resistant bacteria like \textit{Staphylococcus aureus} \cite{plumet2025phage} and \textit{Pseudomonas aeruginosa} \cite{kovacs2024combinations}. Beyond clinical medicine, phages are increasingly used as biocontrol agents in the food industry. Applied to food products or processing surfaces, they can selectively remove contaminants such as \textit{Listeria} and \textit{Salmonella}, thereby enhancing food safety and extending shelf life without affecting the food's quality \cite{bumunang2023bacteriophages}.


The success of these applications critically depends on accurately identifying which phages can infect and eliminate specific bacteria. However, experimental identification of these virus-bacteria interactions is challenging. Standard isolation methods are labor-intensive and, more critically, are limited because phage isolation and cultivation necessitate prior knowledge and culturing of the bacterial host \cite{batinovic2019bacteriophages}. The rapid advancement of high-throughput sequencing has led to an exponential growth in viral sequence data, with metagenomic studies uncovering millions of previously unknown viral genomes. Consequently, the field has increasingly turned to computational approaches to predict phage-host interactions directly from sequence data. Early methods usually relied on identifying virus-host interaction signals like integrated prophages and CRISPR spacers or signatures such as similarities in $k$-mer frequencies or codon usage. While these methods can obtain some reliable results, no single approach is universally effective, as each feature has inherent limitations. For example, alignment-based methods require sufficiently close relatives in databases to make a match, and CRISPR-based predictions are only possible for hosts that have previously recorded a specific phage infection. These limitations have motivated the development of more sophisticated learning models that integrate genomic features to improve predictive accuracy. However, the performance of these models is constrained by three significant and interconnected challenges.

First, a substantial annotation gap exists within the vast amount of sequence data. The number of viral sequences in public databases has grown exponentially, with recent estimates showing a nearly five-fold increase in phage-like assemblies in the GenBank database over the last decade (16,232 in 2015 vs. 78,002 in 2025). However, a large fraction of these sequences represent viral ``dark matter'' with no known host. Even within the more stringently curated RefSeq database, only around 87\% (4,698/5,371 in 2025) viruses have precise host annotation. Furthermore, such annotations lack the required granularity, providing neither the specific host genome nor details about the nature of the virus-host relationship. As illustrated in Figure \ref{fig:intro}, a phage-host interaction is a multi-stage process that includes initial binding (both host or non-host), DNA injection, and either a productive infection (lytic or lysogenic cycle) or failure due to prokaryotic defense systems. A simple taxonomy name of the ``host'' does not distinguish between a phage that can only attach to a cell's surface and one that can successfully replicate within it. Consequently, a model trained on such data may learn to predict successful binding rather than a productive infection, leading to functionally misleading results. Second, the available host annotation data is heavily skewed by historical research biases. A disproportionate number of phages in our databases are linked to a small set of well-studied model organisms, such as \textit{Escherichia coli}, \textit{Salmonella enterica}, and \textit{Mycolicibacterium smegmatis}. This creates a long-tail distribution, where a few bacterial species are associated with thousands of phages, while the vast majority of bacteria have few or no known viral predators in the databases. Such imbalance can introduce significant bias into models, leading to prediction tools that perform well for common hosts but fail to generalize to the broader, more sparsely populated bacterial domain. Finally, the inherent complexity of host range itself presents a major hurdle. Database annotations often imply a simple one-to-one relationship between a phage and a single host species. This administrative simplification cannot reflect the biological reality. Recent studies show that many phages have a relatively broad host range, capable of infecting multiple species or even crossing genus boundaries \cite{ross2016more, chung2023bacteriophages}. For instance, some phages are known to infect multiple distinct species within the \textit{Enterobacteriaceae} family. This broad-host-range behavior is difficult to capture and predict with models trained on simplified, single-host labels, complicating the design of truly comprehensive prediction tools. These combined challenges highlight the difficulty of host prediction and underscore the need for robust computational approaches to navigate this complex data.

The urgent need for virus-host prediction methods has spurred the development of a multitude of computational tools. This rapid growth has created a complex and often confusing landscape where most tools are evaluated on disparate datasets, making direct performance comparisons difficult. While previous review \cite{howell2024computational} provided a valuable initial assessment, its evaluation was limited to a small subset of available tools and relied on a benchmark using only three groups of bacteriophages. Here, we address these gaps by establishing a clear problem framework, providing a systematic analysis of the biological features used by 27 existing tools, and conducting a rigorous benchmark designed to serve as a new standard for the field.

By providing a rigorous, comparative analysis, this review serves two critical functions. First, it acts as a practical guide for researchers selecting the optimal tool for their work. Second, it establishes a foundational resource and a new performance benchmark intended to drive the next wave of innovation in deciphering the complex web of virus-host interactions.

Our key contributions are:

\begin{itemize}
    \item \textbf{A standardized problem formulation:} We establish a structured framework for the virus-host prediction problem, providing a consistent foundation for understanding and comparing diverse computational strategies.
    \vspace{-0.1cm}
    \item \textbf{A comprehensive survey of tools and biological features:} We critically evaluate 27 existing tools, examining their methodologies, strengths, and weaknesses. We then survey the full spectrum of biological features they employed, from CRISPR spacer matching and prophage detection to alignment-free $k$-mer frequency analysis.
    \vspace{-0.1cm}
    \item \textbf{Rigorous benchmarking using carefully designed datasets:} Another key novelty is the development and application of two distinct evaluation benchmarks. \textbf{RefSeq-VHDB} provides a curated set of phage-host pairs for standardized assessment, while \textbf{MetaHiC-VHDB} consists of three independent metagenomic Hi-C test sets designed to assess tool performance in realistic ecological contexts. These benchmarks provide a practical guide for researchers and expose performance gaps that future methods must address.
\end{itemize}

\begin{figure*}
    \centering
    \vspace{-0.5cm}
    \includegraphics[width=0.8\linewidth]{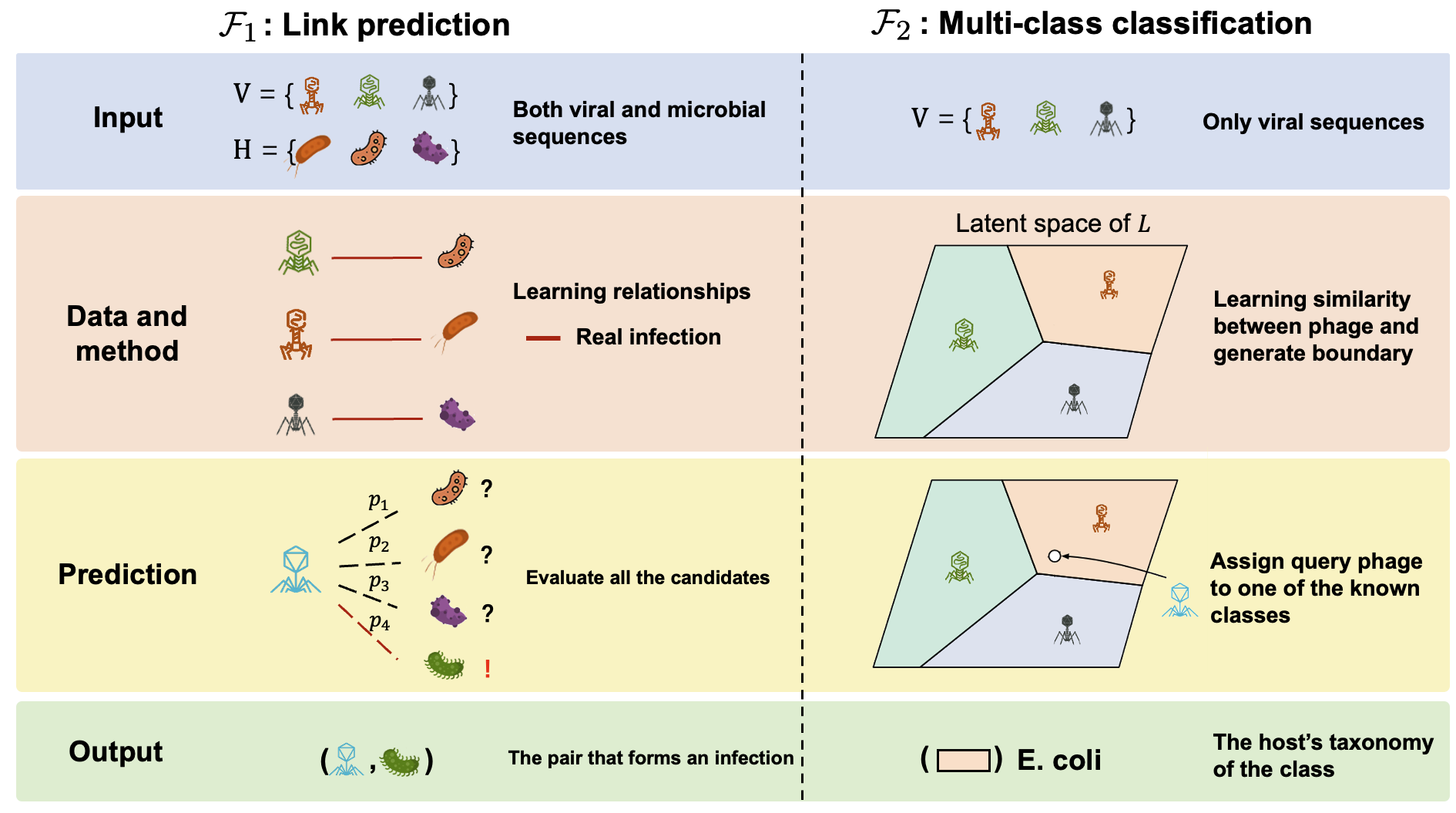}
    \caption{Two primary problem formulations for virus-host prediction task. $\mathcal{F}_1$: Link prediction. $\mathcal{F}_2$: multi-class classification. Based on the input. If a tool uses both viral and microbial sequences as inputs, it can formulate as $\mathcal{F}_1$. To make a prediction for a new virus, it evaluates all potential host candidates to identify the most likely interacting pair. The primary strength of this approach is its generality, making it suitable for discovering novel interactions (de novo discovery) as it is not limited to a known set of hosts. However, it is computationally intensive and must overcome the challenge of extreme class imbalance, where non-interacting pairs vastly outnumber true interactions. If a tool only relies on viruses for host prediction, it usually formulates $\mathcal{F}_2$, aims to separate viruses based on their known hosts, effectively generating a boundary for each class. This method is computationally efficient and precise for hosts represented in the training data. Its main weakness is the closed-set assumption, which prevents it from predicting hosts from taxa that were not included in the training set. }
    \vspace{-0.2cm}
    \label{fig:formulation}
\end{figure*}

\section{Results}
\subsection{Problem formulation of the virus-host prediction}
The structure of publicly available databases, which predominantly document one-to-one virus-host interactions, has directly affected the development of computational tools. Consequently, most methods are designed to predict a single, highest-ranking host for a given virus. This has led to two primary problem formulations for host prediction (Fig. \ref{fig:formulation}). Consider a microbial community containing $N$ viruses ($\{ v_1, \ldots, v_N \}$) and $M$ prokaryotes ($\{h_1, \ldots, h_M\}$). The first and more intuitive formulation frames the task as a link prediction problem on all potential $N \times M$ interactions. The objective is to learn a scoring function $\mathcal{F}_1$ (Eqn. \ref{Eq:f1}). 

\vspace{-0.5cm}
\begin{equation}
\mathcal{F}_1: (v_i, h_j) \mapsto p,\ where\ p \in [0,1]
\label{Eq:f1}
\end{equation}

\noindent where $p$ is the predicted probability that the pair $(v_i, h_j)$ represents a true biological infection.

An alternative formulation simplifies the problem by reframing it as multi-class classification. Instead of predicting individual links, this approach groups the $M$ prokaryotes into $L$ distinct taxonomic ranks (e.g., genus or family). The goal is then to learn a classifier $\mathcal{F}_2$ that maps each virus to a host taxon (Eqn. \ref{Eq:f2}).

\vspace{-0.5cm}
\begin{equation}
\mathcal{F}_2: v_i \mapsto l_k,\ where\ l_k \in \{l_1, \ldots,l_L\}
\label{Eq:f2}
\end{equation}

This formulation operates on the hypothesis that viruses with similar genomic features infect hosts of the same or similar taxa. Consequently, $\mathcal{F}_2$ propagate taxonomic labels from a reference set of viruses with known hosts to novel viral sequences. While both formulations aim to solve the same fundamental problem, they differ significantly in their data requirements, underlying features, and the nature of their predictions, each presenting distinct advantages and disadvantages. A basic comparison is listed in Table. \ref{tab:formulation} and detailed explanation can be found in Supplementary Note 1.

\begin{table*}[h!]
\vspace{-0.3cm}
\centering
\caption{Comparison of link prediction and multi-class classification formulations for virus host prediction .}
\begin{tabular}{
  >{\raggedright\arraybackslash}p{0.18\textwidth} 
  >{\raggedright\arraybackslash}p{0.37\textwidth} 
  >{\raggedright\arraybackslash}p{0.35\textwidth} 
}
\toprule
\textbf{Characteristic} & \textbf{Link Prediction Formulation} & \textbf{Multi-class Classification Formulation} \\
\midrule
\textbf{Primary Goal} &
Predicts whether a specific virus interacts with a specific host (binary decision per pair). &
Assigns a host taxonomic label to a given virus from a predefined set of taxa. \\
\addlinespace 
\textbf{Data Challenge} &
Extreme class imbalance; the number of negative (non-interacting) pairs vastly overwhelms positive pairs. &
Long-tail distribution of class labels; a few host taxa are heavily over-represented, while most are rare. \\
\addlinespace
\textbf{Feature Representation} &
Primarily \textbf{pairwise features} representing direct evidence (e.g., sequence homology, CRISPR-spacer matches, prophages). &
Primarily \textbf{virus-centric features} creating a genomic fingerprint (e.g., marker genes, codon usages, protein organization). \\
\addlinespace
\textbf{Prediction Output} &
An explicit, high-resolution pairing between a viral genome and a host genome. &
A single, lower-resolution taxonomic label (e.g., family or genus) for the host. \\
\addlinespace
\textbf{Key Strength} &
\textbf{Generality}. Well-suited for \textit{de novo} discovery of novel interactions, as it is not constrained by a known set of hosts. &
\textbf{Precision \& Efficiency}. Computationally fast (one prediction per virus) and often highly accurate for hosts within databases. \\
\addlinespace
\textbf{Key Weakness} &
High risk of false positives due to the massive number of potential interactions (number of viruses $\times$ number of prokaryotes). Computationally intensive. &
\textbf{Closed-set assumption}; cannot predict hosts from taxa absent in the training data. Struggles with polyvalent viruses. \\
\bottomrule
\end{tabular}
\label{tab:formulation}
\vspace{-0.2cm}
\end{table*}

\begin{sidewaystable*}[thp]
\centering
\caption{A comparison of models/methods employed for host prediction of prokaryotic viruses. $\mathcal{F}_1$: link-prediction formulation. $\mathcal{F}_2$: multi-class classification formulation. The ``Inputs'' of a tool represent which kinds of data required by the methods: ``Virus'' means the tools only required viral sequences as inputs; ``Virus and MAGs'' means both the viral sequences and a set of candidate host genomes or Metagenome-Assembled Genomes (MAGs) from the same sample are required. ``Either'' means the tool accepts inputs in either way. The ``Prediction level'' of a tool represents the lowest evaluated host taxonomy level reported in the origin paper. The ``Availability'' of a tool from an end-user standpoint was evaluated based on three key criteria: ease of installation, quality of documentation, and workflow automation. A designation of ``Yes'' indicates that the software met all three usability requirements. For tools marked as ``No'', which failed to meet one or more of these criteria, detailed information can be found in the Supplementary Note 1.}
\begin{tabular}{cccccccc}
\hline
 Name & Formulation & Method & Feature & Inputs & Prediction level & Availability & Ref. \\ \hline
\rowcolor{gray!10} HostPhinder (2016) & $\mathcal{F}_2$ & Homology analysis                      & $k$-mers                        & Virus     & Species-level & No & \cite{villarroel2016hostphinder}\\ 
\rowcolor{gray!30} WIsH (2017)        & $\mathcal{F}_1$ & Hidden markov model                    & $k$-mers                        & Virus and MAGs     & Genus-level & Yes & \cite{galiez2017wish}\\ 
\rowcolor{gray!30} VHM (2017)         & $\mathcal{F}_1$ & Homology analysis                      & $k$-mers                        & Virus and MAGs     & Species-level & Yes & \cite{ahlgren2017alignment}\\ 
\rowcolor{gray!30} PHP (2021)         & $\mathcal{F}_1$ & Gaussian Mixture Models                & $k$-mers                        & Either     & Genus-level & Yes &  \cite{lu2021prokaryotic}\\
\rowcolor{gray!10} HostG (2021)       & $\mathcal{F}_2$ & Graph convolutional network            & Integrated multiple features    & Virus     & Genus-level & No & \cite{shang2021predicting}\\ 
\rowcolor{gray!10} RaFAH (2021)       & $\mathcal{F}_2$ & Random forest model                    & Viral protein similarity        & Virus     & Genus-level & Yes & \cite{coutinho2021rafah}\\ 
\rowcolor{gray!10} VPF-Class (2021)   & $\mathcal{F}_2$ & Homology analysis                      & Viral protein similarity        & Virus     & Genus-level & No & \cite{pons2021vpf}\\ 
\rowcolor{gray!30} PredPHI (2022)     & $\mathcal{F}_1$ & Convolutional neural network           & Viral protein similarity        & Virus and MAGs     & Genus-level & No & \cite{li2020deep}\\ 
\rowcolor{gray!30} VHM-Net (2022)     & $\mathcal{F}_2$ & Hybrid model                           & Integrated multiple features    & Either     & Species-level & Yes & \cite{wang2020network}\\ 
\rowcolor{gray!30} PHIAF (2022)       & $\mathcal{F}_1$ & Convolutional neural network           & $k$-mers                        & Virus and MAGs     & Species-level & No & \cite{li2022phiaf}\\ 
\rowcolor{gray!30} PHIST (2022)       & $\mathcal{F}_1$ & Homology analysis                      & $k$-mers                        & Virus and MAGs     & Species-level & Yes & \cite{zielezinski2022phist}\\ 
\rowcolor{gray!30} CHERRY (2022)      & $\mathcal{F}_1$ & Hybrid model              & Integrated multiple features    & Either     & Species-level & Yes & \cite{shang2022cherry}\\ 
\rowcolor{gray!10} DeepHost (2022)    & $\mathcal{F}_2$ & Convolutional Neural Network           & $k$-mers                        & Virus      & Species-level & Yes & \cite{ruohan2022deephost}\\ 
\rowcolor{gray!10} HoPhage (2022)     & $\mathcal{F}_2$ & Hybrid model                           & Viral protein similarity        & Virus      & Genus-level & No & \cite{tan2022hophage}\\ 
\rowcolor{gray!10} vHULK (2022)       & $\mathcal{F}_2$ & deep neural network                   & Viral protein similarity        & Virus      & Species-level & Yes & \cite{amgarten2022vhulk}\\ 
\rowcolor{gray!30} iPHoP (2023)       & $\mathcal{F}_1$ & Hybrid model                           & Integrated multiple features    & Either     & Species-level & Yes & \cite{roux2023iphop}\\ 
\rowcolor{gray!30} PhageTB (2023)     & $\mathcal{F}_1$ & Hybrid model                           & Integrated multiple features    & Virus and MAGs     & Genus-level & No & \cite{aggarwal2023ensemble}\\ 
\rowcolor{gray!30} PHPGCA (2023)      & $\mathcal{F}_1$ & Light Graph Convolutional Network      & Integrated multiple features    & Virus and MAGs     & Species-level & No & \cite{du2023prokaryotic}\\ 
\rowcolor{gray!10} PHERI (2023)       & $\mathcal{F}_2$ & Decision tree                          & Viral protein similarity        & Virus      & Genus-level & Yes & \cite{balavz2023pheri}\\ 
\rowcolor{gray!10} PHIEmbed (2023)    & $\mathcal{F}_2$ & Random forest model                    & Viral marker protein            & Virus      & Genus-level & No & \cite{gonzales2023protein}\\
\rowcolor{gray!30} DeepPBI-KG (2024)  & $\mathcal{F}_1$ & Deep neural network           & Sequence similarity             & Virus and MAGs      & Species-level & No & \cite{wei2024deeppbi}\\
\rowcolor{gray!30} VHIP (2024)        & $\mathcal{F}_1$ & Gradient boosting classifier                         & $k$-mers                        & Virus and MAGs     & Species-level & No & \cite{bastien2024virus}\\
\rowcolor{gray!30} PB-LKS (2024)      & $\mathcal{F}_1$ & Homology analysis                      & $k$-mers                        & Virus and MAGs     & Genus-level & Yes & \cite{qiu2024pb}\\
\rowcolor{gray!10} EvoMIL (2024)      & $\mathcal{F}_2$ & multiple instance learning model       & Viral protein similarity        & Virus     & Species-level & No & \cite{liu2024prediction}\\
\rowcolor{gray!30} PHPGAT (2025)      & $\mathcal{F}_1$ & Graph Attention Network v2             & Integrated multiple features    & Virus and MAGs     & Species-level & No & \cite{liu2025phpgat}\\
\rowcolor{gray!30} MI-RGC (2025)      & $\mathcal{F}_1$ & Regional graph convolution             & $k$-mers                        & Virus and MAGs     & Species-level & No & \cite{wei2024predicting}\\
\rowcolor{gray!10} PHIStruct (2025)   & $\mathcal{F}_2$ & Protein language model & Viral marker protein            & Virus     & Genus-level & No & \cite{gonzales2025phistruct} \\ \hline
\end{tabular}
\label{tab:summary}
\end{sidewaystable*}

\begin{figure*}
    \centering
    \vspace{-0.3cm}
    \includegraphics[width=1\linewidth]{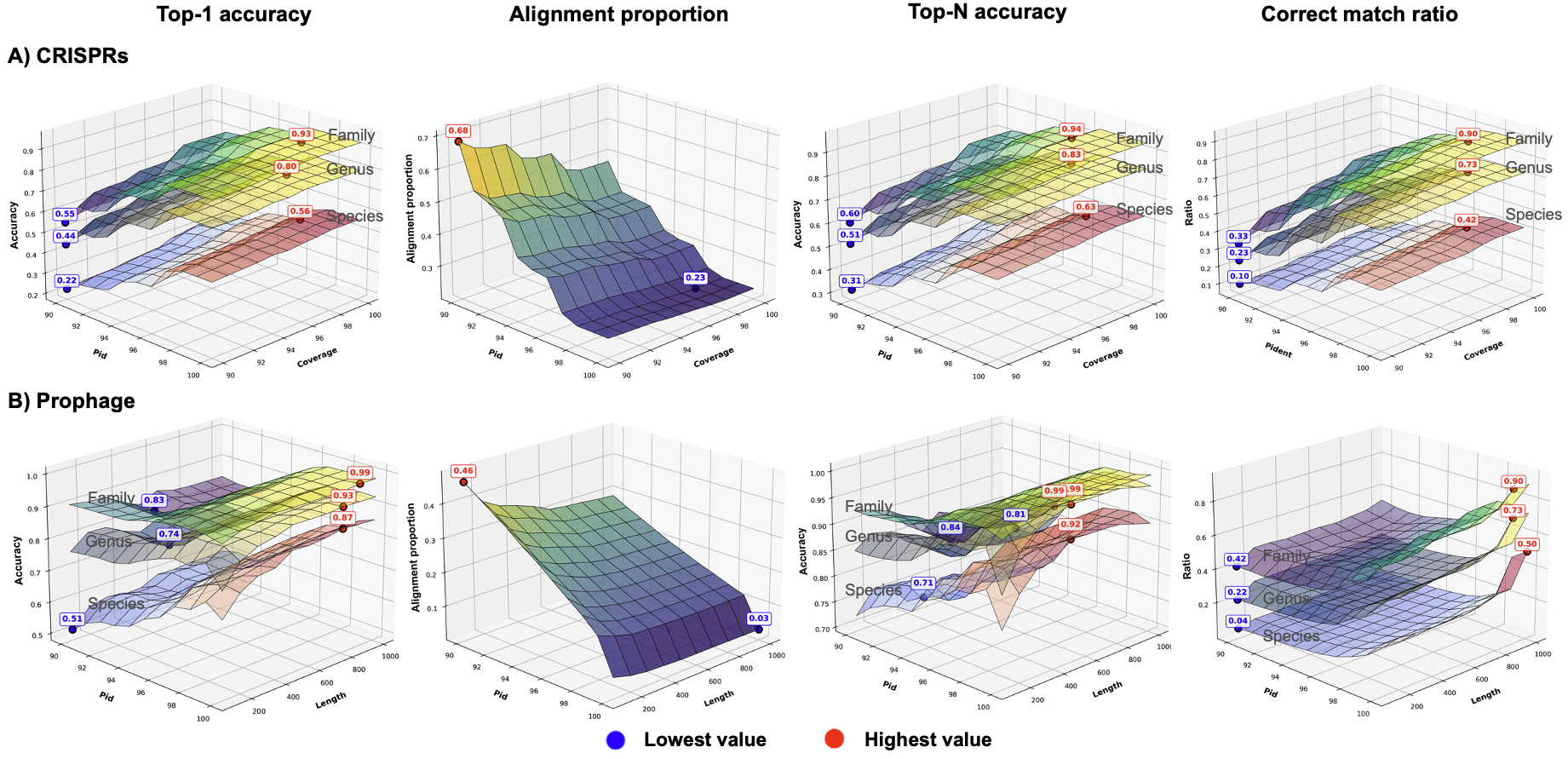}
    \vspace{-0.5cm}
    \caption{Surface plots to illustrate the utility of (A)  CRISPR spacer match and (B)  prophage homologous search. Top-1 accuracy: accuracy summarized on the best hit (the hit with the best alignment score). Performance is evaluated at the Species, Genus, and Family taxonomic levels, based on key alignment parameters. For CRISPRs, these parameters are percentage of identical matches (pid) and coverage, while for prophages, they are pid and alignment length. The lowest and highest values on each surface are highlighted with blue and red dots, respectively. The four metrics evaluated are: Top-N accuracy: accuracy summarized on the all the alignment hits. Alignment proportion: the fraction of viruses with at least one alignments. Correct match ratio: a proportion of the correct hits in the alignment results, estimating the errors introduced by using Top-N accuracy. }
    \vspace{-0.2cm}
    \label{fig:alignment}
\end{figure*}

\subsection{Current approaches for host prediction}
We thoroughly reviewed 27 computational methods for host prediction, which we group by their problem formulation: link prediction or multi-class classification. Our chronological analysis reveals a distinct technological evolution in the features and algorithms used. A brief summary of these tools is listed in Table \ref{tab:summary}. Early methods, such as WIsH and VirHostMatcher (VHM), primarily relied on alignment-free genomic signatures like $k$-mer and oligonucleotide frequencies. The field then progressed toward integrating multiple biological signals, with tools like VirHostMatcher-Net (VHM-Net) and iPHoP combining CRISPR matches, sequence homology, and the outputs of other predictors into more robust frameworks. This trend was followed by a rapid adoption of deep learning, beginning with Convolutional Neural Networks (CNNs) that learned from sequence-based features (e.g., DeepHost, PHIAF). More recently, the research has advanced to sophisticated graph-based deep learning models (e.g., CHERRY, PHPGAT) that capture complex relationships within heterogeneous biological networks. The latest innovations leverage large protein language models to generate rich embeddings from viral proteins, particularly receptor-binding proteins, for highly targeted predictions (e.g., EvoMIL, PHIStruct). Regarding the problem formulation, tools based on link prediction require both viral and host sequences as input, whereas those based on multi-class classification rely only on viral sequences. Notably, although PHP, iPHoP, and CHERRY are designed using a link prediction framework, they incorporate internal, pre-compiled databases. This design enables them to generate predictions using only viral sequences as input. Despite this algorithmic progress, our practical assessment highlights a critical challenge: over half of the reviewed tools are not readily usable due to issues such as failed installations, hard-coded dependencies, or a lack of documentation, severely limiting their reproducibility and broader utility (see Supplementary Note 2).

\subsection{Genomic features vary in their ability to predict hosts}
The predictive power of computational models for host prediction hinges on the biological features extracted from viral genomic sequences. These features broadly fall into two categories: signals of potential virus-host interaction and signals of virus-virus similarity. Here, we systematically reviewed the utility of the most common genomic features, including CRISPR-Cas spacer matches, prophage detection, similarities in $k$-mer frequencies, and viral DNA/protein sequence homology. To establish a labeled data for our analysis, we utilized a dataset of 4,698 viruses with species-level host annotations from the Viral RefSeq database. For the host genomic data, we downloaded all 110,988 available prokaryotic genomes from GenBank. We then evaluated the effectiveness of each genomic feature in predicting viral hosts, thereby delineating their individual strengths and limitations.


\subsubsection{CRISPR spacer match}
CRISPR-Cas systems provide prokaryotes with an adaptive immune record of past viral infections by integrating short viral DNA fragments (spacers) into the host genome. A match between a viral sequence and a CRISPR spacer is therefore high-confidence evidence of an interaction, making it a cornerstone in many virus-host interaction analysis \cite{nayfach2021metagenomic,johansen2023centenarians}. To evaluate the predictive utility of CRISPR spacer match, we aligned the genomes of 4,698 viruses against 2,005,489 CRISPR spacers identified from 110,988 prokaryotic genomes in GenBank (see Methods). Our analysis revealed that while 48\% of prokaryotic genomes contain at least one CRISPR array (Supplementary Fig. S1), only a fraction of them yield viral matches. Overall, spacer alignments successfully linked 76\% of the viruses to 12\% of the host genomes in the dataset. Analysis of the top ten phyla with the most spacers found revealed heterogeneity in both the prevalence of CRISPR systems and the spacer hits (Supplementary Fig. S1). We observed a clear disconnect between the sheer quantity of spacers and the rate of successful virus-host linkage. For instance, the phylum \textit{Thermodesulfobacteriota} contains the largest number of genomes and the highest absolute count of CRISPR spacers; however, phyla like \textit{Spirochaetota} and \textit{Fusobacteriota} have a larger percentage of genomes with a spacer match.

Then, we performed a grid search on percentage of identical matches (pid) and alignment coverage (defined as the alignment length divided by the spacer length), with both parameters starting at 90\%. These thresholds are widely used in many research to find CRISPR spacer matches \cite{nayfach2021metagenomic,johansen2023centenarians}. The performance was quantified using four metrics (see Methods): Top-1 accuracy (accuracy on the best hit), Top-N accuracy (accuracy on all alignment), alignment proportion (proportion of aligned viruses), and correct match ratio (estimate the errors introduced by using Top-N accuracy). Fig. \ref{fig:alignment}A reveals that prediction accuracy is highly sensitive to the percentage of identical matches (pid) but less sensitive to alignment coverage. While stringent thresholds (e.g., pid $\geq$ 98\%, coverage $\geq$ 96\%) are required for reliable predictions, this stringency significantly reduces the proportion of viruses for which a host can be found. At optimal thresholds, Top-1 accuracy reached 57\% at the species level and 82\% at the genus level. The higher genus-level accuracy may arise because a successful infection implies a failure of the host's CRISPR system, whereas other species in the same genus may have successfully neutralized the virus and recorded the spacer. Furthermore, our analysis indicates that considering all alignments as predictions leads to ambiguous results (low correct match ratio) while providing only a marginal increase in Top-N accuracy. Thus, while highly reliable, CRISPR-based predictions are sparse, necessitating complementary features.

\begin{figure*}
    \vspace{-0.2cm}
    \centering
    \includegraphics[width=0.92\linewidth]{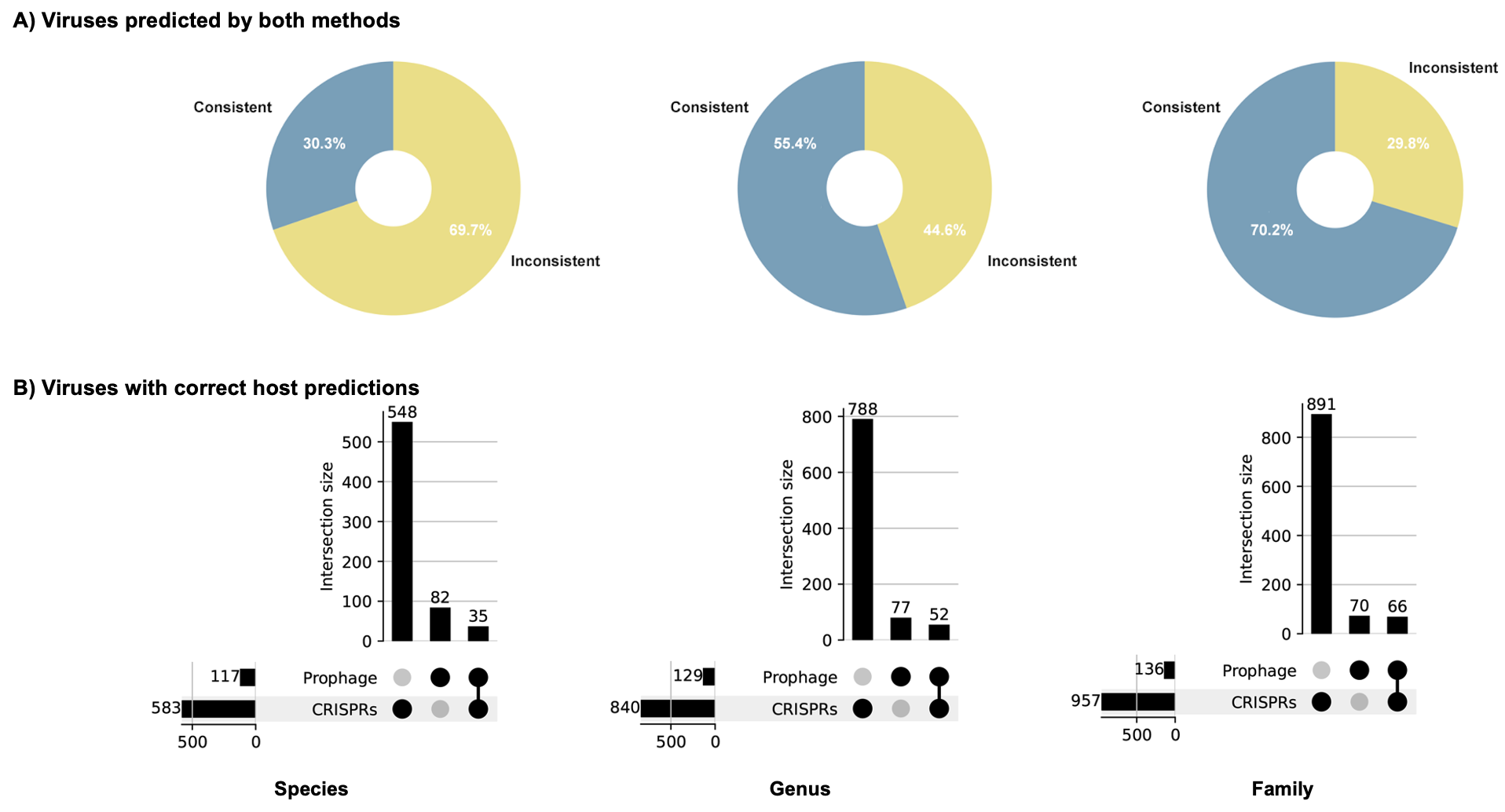}
    \vspace{-0.2cm}
    \caption{Complementarity of CRISPR-based and prophage-based virus-host predictions. (A) Consistency of host predictions for viruses identified by both methods. The charts show the percentage of predictions that are consistent (blue) versus inconsistent (yellow) at the species, genus, and family levels. (B) Overlap of correct virus-host predictions from CRISPR spacer analysis and prophage homology searching. The UpSet plots show the number of correct predictions unique to each method (single dots for prophage unique and CRISPRs unique) and the number shared by both (connected dots) across the three taxonomic ranks. The horizontal bars indicate the total number of correct predictions for each method.}
    \label{fig:consist}
    \vspace{-0.2cm}
\end{figure*}

\subsubsection{Prophage matches}
The integration of a viral genome into a host chromosome as a prophage is clear evidence of a lysogenic relationship. This signal is typically identified by detecting prophage regions in microbial genomes or by direct DNA alignment between viral and microbial genomes. To illustrate the utility of this feature, we conducted the most commonly used BLASTN-based alignment search between viral and prokaryotic genomes (see Methods). 

According to the alignment results, 59\% of prokaryotic genomes and 51\% of viruses are linked with at least one alignment (Supplementary Fig. S2). Meanwhile the order Top-10 phyla with the most BLASTN alignments is significantly different compared to the one in CRISPR spacer matches (Supplementary Fig. S2), demonstrating the natural bias introduced by these features. A plausible explanation is that the existence of prophages indicates infection, which the CRISPR system fails to prevent. Following a similar methodology to the CRISPR alignment experiment, we evaluated performance across a grid search of two widely used threshold combination: pid, starting from 90\% and alignment length (starting from 100 bp). The BLASTN-based alignment reveals that prophage searches yield higher species-level accuracy than CRISPR analysis but suffer from a lower alignment proportion and more ambiguous hits at relaxed thresholds (Fig. \ref{fig:alignment}B). A pid of 95\% and an alignment length $\geq$ 500 bp provides a favorable balance for Top-1 accuracy and alignment proportion. However, when using all alignments (Top-N) to include more candidate predictions, a higher pid (e.g., $\geq$ 98\%) should be considered to minimize false positives.

\begin{figure*}
    \centering
    \includegraphics[width=0.9\linewidth]{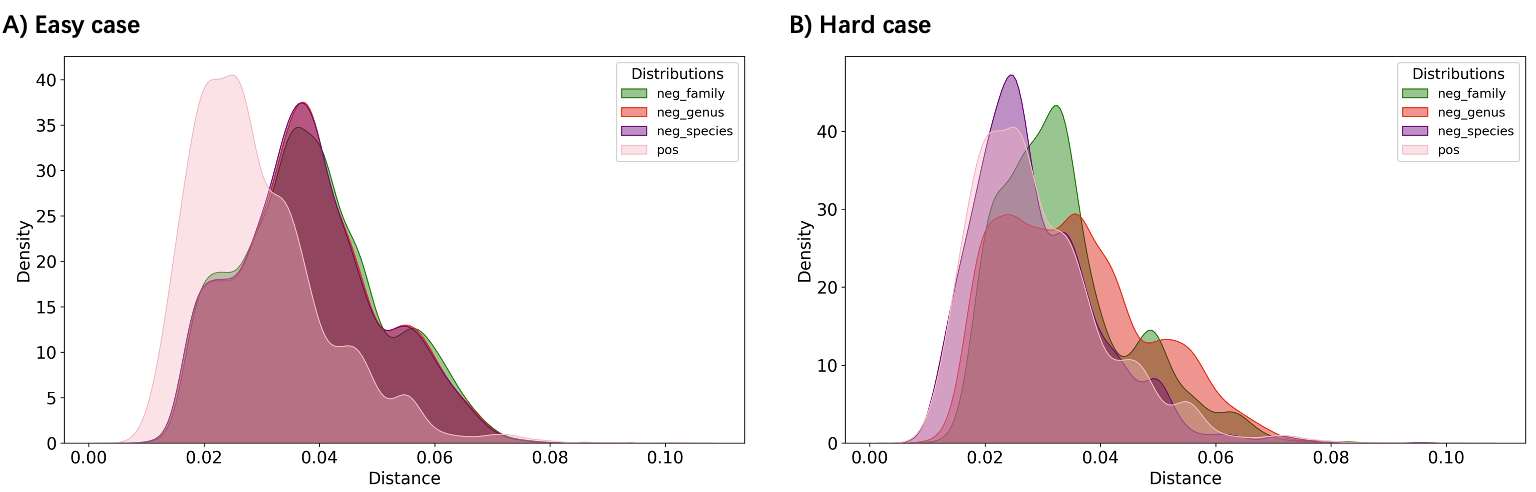}
    \caption{4-mer frequency distance distribution of known virus-host pairs (positive pairs) from Genbank against those of non-host pairs (negative pairs). A: The ``easy'' negative case includes negative pairs where the prokaryotes are in different taxa from the given host (e.g., species-level taxonomy). B: The ``hard'' negative case has a constraint on the highest taxonomy rank of the prokaryotes (e.g., prokaryotes do not belong to the same species but belong to the same genus as the host). }
    \label{fig:kmer}
\end{figure*}

\subsubsection{Consistency of CRISPR spacer match and prophage detection} As aforementioned, CRISPR spacer match and prophage detection introduce a significant bias on the phyla of prokaryotes (Supplementary Fig. S1 and Fig. S2). Here, we estimate the potential of using them as complementary for host prediction. First, we conduct a comparative analysis of the host predictions for a cohort of viruses that have annotations from both methodologies. Fig. \ref{fig:consist}A illustrates that the consistency for viruses annotated by both methods was low at the species level (30.3\%) but increased at higher taxonomic ranks (55.4\% at genus, 70.2\% at family). We further examined the overlap of correctly predicted virus-host pairs between the two methods. The UpSet plot (Fig. \ref{fig:consist}B) analysis confirms that the two methods identify largely distinct sets of virus-host interactions. At the species level, only 35 interactions were identified by both methods, whereas CRISPR and prophage searches uniquely identified 548 and 82, respectively. This minimal overlap persists across taxonomic ranks, suggesting that these features can be used in conjunction to maximize the discovery of virus-host linkages.

\subsubsection{$k$-mer frequency similarity}
Viruses frequently adapt their genomic composition, including aspects like codon usage, to resemble that of their hosts. This phenomenon, often termed a 'genomic signature,' can be quantified using $k$-mer frequency profiles. Consequently, host prediction can be performed by measuring the distance (e.g., cosine similarity, Euclidean distance) between the $k$-mer frequency vectors of a virus and those of potential hosts.

To assess its discriminative power, we compared the $k$-mer frequency distributions between known virus-host pairs and non-host pairs (see Methods). We chose $k = 4$ because it is the most commonly used setting and offers a balance between informational content and computational cost. In the analysis, we defined two types of negative pairs to simulate different challenge levels. An example is illustrated in Supplementary Fig. S3. The ``easy'' negative case includes negative pairs where the prokaryotes are in different taxa from the given host, and the ``hard'' negative case has a constraint on the taxonomy rank of the chosen prokaryotes (e.g., prokaryotes do not belong to the same species but belong to the same genus as the host).

The results reveal that $k$-mer profiles effectively distinguish viruses from distantly related non-hosts (the "easy" negative case), showing a clear bimodal distribution of distances (Fig. \ref{fig:kmer}A). However, the distributions for hosts and other prokaryotes within the same host genus (the "hard" negative case) overlap considerably, limiting the feature's resolving power at the species level (Fig. \ref{fig:kmer}B). 
This highlights a critical challenge for building models: training datasets built on "easy" negatives may distinguish trivial cases but fail to generalize to high-resolution, intra-genus predictions.

\subsubsection{Host-specific marker genes}
Host-specific marker genes, such as Receptor-Binding Proteins (RBPs) in tailed phages, offer direct mechanistic links to host specificity. Homology between RBPs can strongly suggest a shared host range. However, the utility of this feature is limited. First, it is primarily applicable to specific viral clades like \textit{Caudoviricetes}, as RBPs are not universally conserved. Second, high sequence diversity makes RBP identification computationally challenging, with even state-of-the-art tools achieving an F1-score of only around 0.8 \cite{boeckaerts2022identification}, thereby hindering their broad application in host prediction.

\begin{figure}[ht]
    \centering
    \includegraphics[width=0.5\linewidth]{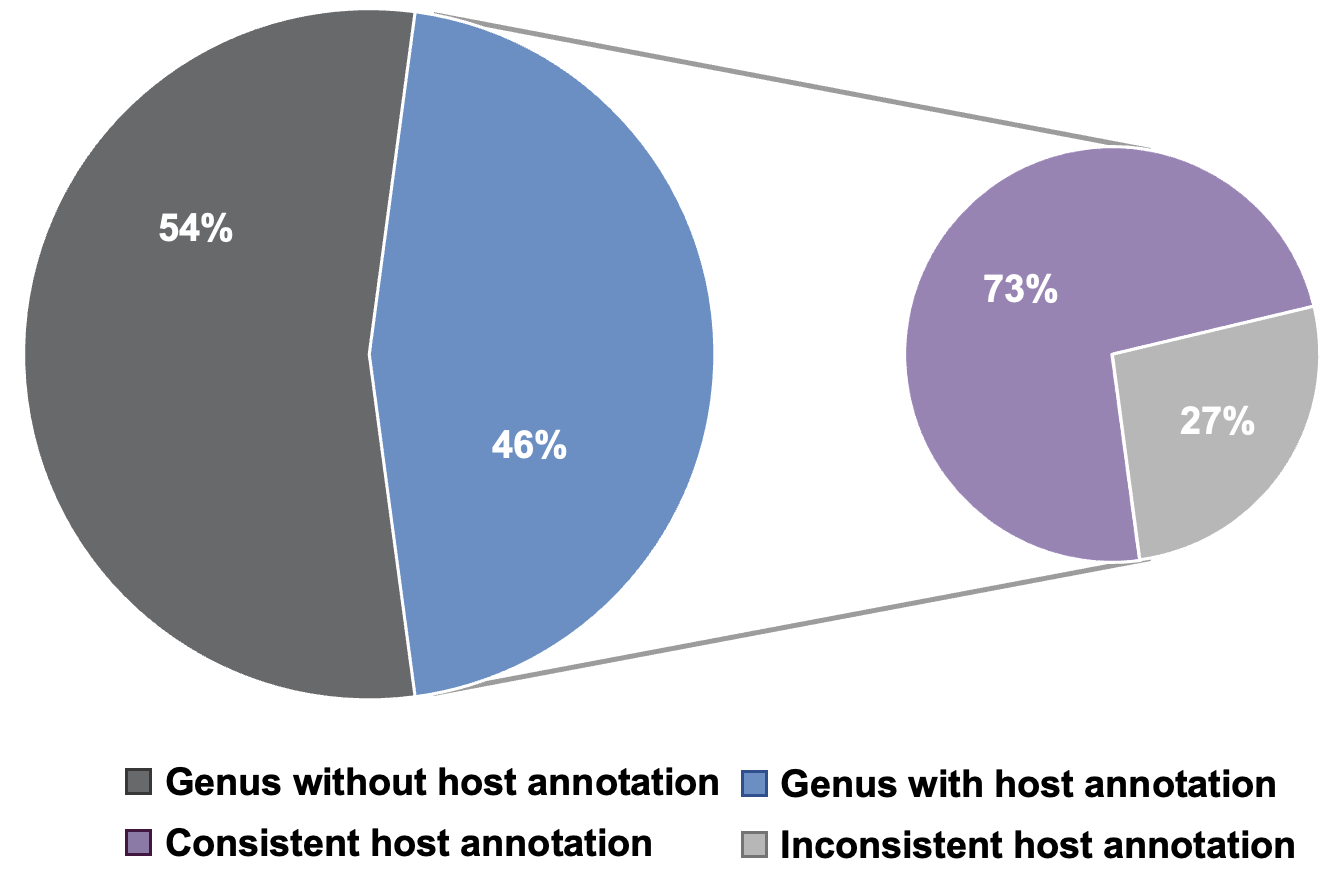}
    \caption{Proportion of viral genera with consistent host annotations in the Viral RefSeq database. The pie chart on the left shows that a majority of genera (54\%) lack any host annotation, while the remaining 46\% (1,263 genera) are annotated. The chart on the right further examines this annotated portion, revealing that 73\% (927 genera) have a consistent host annotation.}
    \label{fig:proportion}
\end{figure}

\subsubsection{Viral genomic similarity}
A more general approach relies on viral genomic similarity, measured by shared protein content, synteny, or $k$-mer frequency, between a query virus and reference viruses in the database. This method is founded on the observation that viruses belonging to the same taxonomic group (e.g., genus) often infect a similar hosts. To validate this assumption, we analyzed the Viral RefSeq database. Our results confirmed that this principle holds for a significant subset of viruses: 73\% of genera (927 of 1,263) with species-level host annotations exhibit perfect host consistency, defined as all viruses within the genus infecting the same host species (Fig. \ref{fig:proportion}). However, only 45\% of the genera (1,263 of 2,760) have clear host annotations. Consequently, the practical utility of this method is constrained by the current state of database annotation. The prevalence of viruses with unknown hosts or incomplete taxonomic information reduces the size and resolution of the reference set, thereby limiting the predictive power of this approach.

\subsection{Benchmarking reveals a trade-off between performance and efficiency in host prediction}
A comprehensive evaluation of existing virus-host prediction methods is essential for providing practical guidance to researchers. In practice, two primary application scenarios for these methods can be defined. The first scenario involves the large-scale analysis of viral sequences, often sourced from public databases, where corresponding host genomes are unavailable. In this context, predictions must be made using only the viral sequences as input. The second scenario involves the analysis of specific metagenomic datasets that contain both viral sequences and a set of candidate host genomes or Metagenome-Assembled Genomes (MAGs) from the same sample. Here, prediction tools can leverage information from both the virus and potential hosts. To evaluate tool performance in both contexts, we constructed two distinct benchmark datasets: RefSeq-VHDB and MetaHiC-VHDB, each designed to simulate one of these scenarios (see Methods). Then, we present a performance comparison of the available tools previously identified in Table \ref{tab:summary}, evaluating them on both benchmark datasets.

\subsubsection{Benchmarking tools using viral sequence-only data}\label{exp1}
The RefSeq-VHDB dataset contains only viral genomic sequences and lacks corresponding host genomes for each entry. The data property is illustrated in Supplementary Fig. S5. This structural limitation precludes the use of tools that require a user-supplied set of candidate host genomes (i.e., those based on a link prediction framework). Consequently, our comparative analysis on this dataset was restricted to tools capable of predicting a host from the viral sequence alone, including PHP, VHM-Net, RaFAH, DeepHost, vHULK, CHERRY, iPHoP, and PHERI. Notably, while PHP, iPHoP, and CHERRY are fundamentally designed for link prediction, they circumvent this limitation by incorporating their own internal, pre-compiled host databases. This feature enables them to generate predictions using only a viral sequence as input.

\begin{figure*}[!ht]
    \centering
    \vspace{-0.5cm}
    \includegraphics[width=0.95\linewidth]{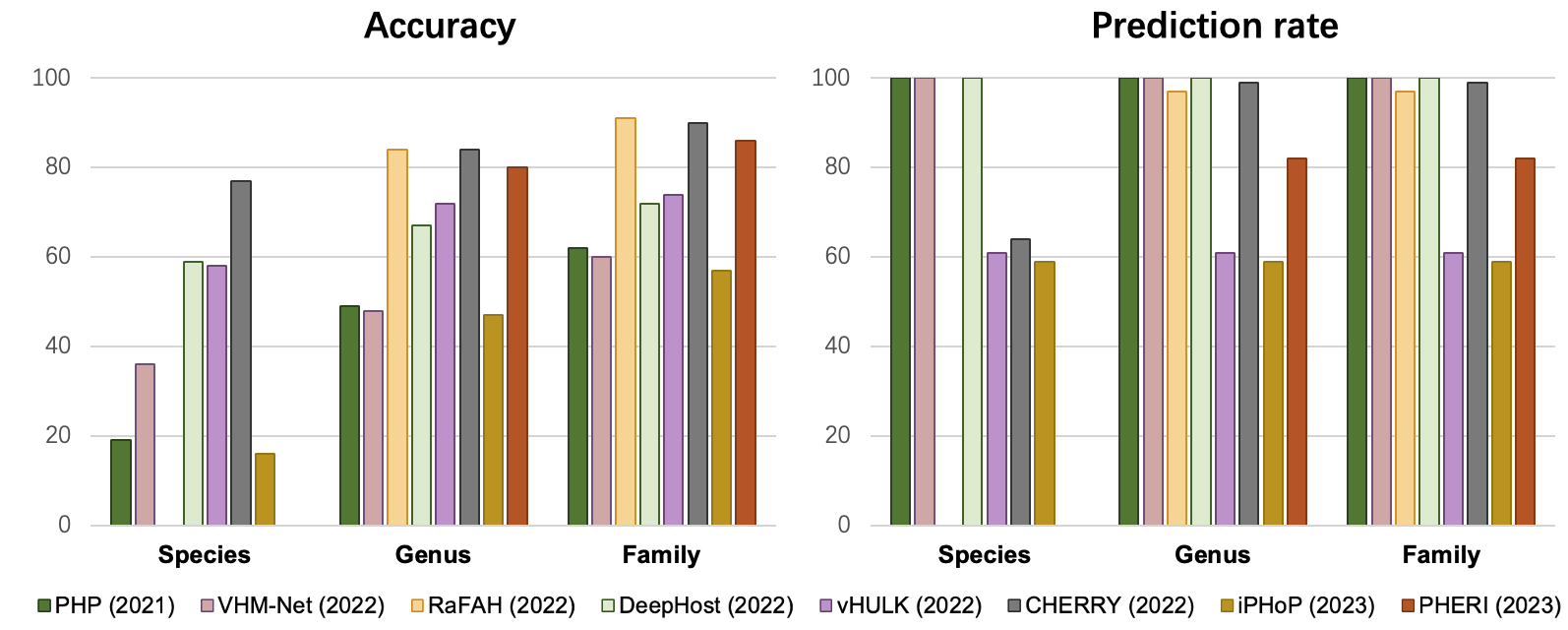}
    \caption{Performance of eight viral host prediction tools on the RefSeq-VHDB benchmark. The tools are evaluated based on prediction accuracy (left panel) and prediction rate (right panel) at the species, genus, and family taxonomic ranks. For tools capable of returning multiple predictions, only the top-ranked result was considered. RaFAH and PHERI do not provide species-level predictions and are therefore omitted from that category. }
    \vspace{-0.5cm}
    \label{fig:refseq}
\end{figure*}

The performance of the eight selected tools on the RefSeq-VHDB benchmark is summarized in Fig. \ref{fig:refseq}. We evaluated each tool based on its prediction accuracy and prediction rate at the species, genus, and family taxonomic ranks (see Methods). For tools capable of returning multiple host predictions for a single virus (e.g., PHP, CHERRY, and iPHoP), we only considered the top-ranked prediction to ensure a fair and standardized comparison. It should be noted that RaFAH and PHERI do not provide species-level predictions; therefore, their performance is reported only at the genus and family levels. Conversely, while PHP was not originally evaluated at the species rank in its publication, the software does produce species-level outputs, which we have included in our analysis.

In terms of prediction accuracy (Fig. \ref{fig:refseq}), a general trend was observed across all methods: performance improved as the taxonomic classification became higher, from species to family. At the species level, CHERRY demonstrated markedly superior performance, correctly identifying the host for approximately 77\% of the predicted viruses. At the genus level, RaFAH achieved the highest accuracy at approximately 84\%, with CHERRY (83\%) and PHERI (80\%) also showing strong performance. This trend continued at the family rank, where RaFAH (92\%) and CHERRY (90\%) were the top performers. These findings lead to a clear set of recommendations. 

The prediction rate analysis reveals two distinct behaviors among the tools (Fig. \ref{fig:refseq}). A subset of the tools—PHP, VHM-Net, and DeepHost—achieved a 100\% prediction rate across all taxonomy ranks, providing a host prediction for every viral sequence in the dataset. In contrast, vHULK, CHERRY, iPHoP and PHERI appear to employ internal confidence thresholds, as they did not return a prediction for every query. This indicates a potential trade-off, where these tools may sacrifice comprehensive prediction rate to maintain a higher certainty for the predictions they do make. For general-purpose applications requiring high accuracy, our analysis identifies CHERRY as the most effective and robust tool, achieving the highest species-level accuracy and top-tier performance at higher taxonomic ranks with a 100\% prediction rate. While lacking species-level resolution, RaFAH is also an excellent choice for genus- and family-level predictions. 

Interestingly, despite arranging the tools chronologically by their publication date on the x-axis, our findings reveal no strong correlation between a tool's release date and its performance. The sustained high performance of certain tools can likely be attributed to their underlying algorithm and choice of biological features. In support of this, we observed that the top-performing tools, as illustrated in Table \ref{tab:summary}, all incorporate features based on viral genomic or protein similarity, demonstrating the critical importance of this approach in this application scenario.

\subsubsection{Benchmarking tools using viruses and MAGs driven from metagenomics}
To evaluate tool performance in a more practical metagenomic context, we next employed a benchmark derived from MetaHiC sequencing data. This dataset, termed MetaHiC-VHDB, comprises 251 virus-host interactions sourced from three distinct environments: 84 from human gut samples \cite{press2017hi}, 66 from bovine fecal samples \cite{stewart2018assembly}, and 101 from wastewater samples \cite{stalder2019linking} (see Methods). In contrast to the RefSeq-VHDB benchmark, this evaluation framework provided tools with both viral contigs and a corresponding set of candidate host MAGs from the same sample. This experimental design is intended to more faithfully replicate the real-world challenge of identifying the specific host of a newly discovered virus within a complex microbial community.

\begin{figure*}
    \centering
    \vspace{-0.2cm}
    \includegraphics[width=0.95\linewidth]{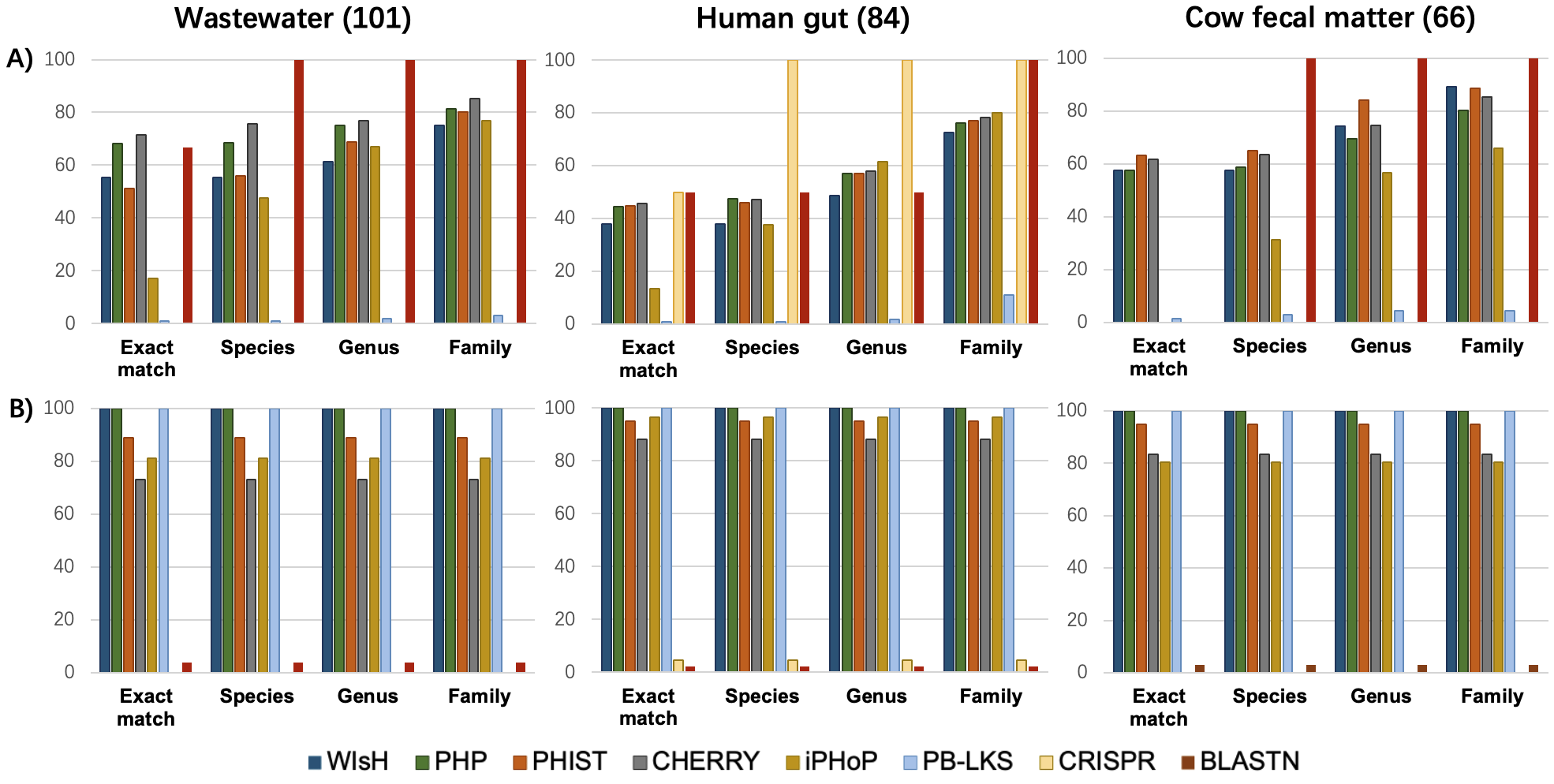}
    \vspace{-0.2cm}
    \caption{Performance comparison of eight viral host prediction tools on three metagenomic datasets. The evaluation was conducted on virus-host pairs from Wastewater, Human gut, and Cow fecal matter samples. The top row (A) displays prediction accuracy, while the bottom row (B) shows the prediction rate. Performance was assessed at four levels: exact match (correct host's genome), species, genus, and family. }
    \label{fig:hic}
    \vspace{-0.2cm}
\end{figure*}

We evaluated the tools compatible with this input format: WIsH, PHP, PHIST, CHERRY, iPHoP, and PB-LKS. Performance was assessed at four taxonomic levels: exact host genome (exact match), species, genus, and family (Fig. \ref{fig:hic}). As expected, the tools generally exhibited improved accuracy at higher taxonomic ranks. Among them, PHP, PHIST, and CHERRY demonstrated the most robust and competitive accuracy across the three datasets. Specifically, PHIST was a top performer at the exact match level in both the cow fecal and human gut datasets. CHERRY exhibited consistent performance across all three environments, but at the cost of a lower prediction rate compared to other tools. While PHP generally achieved slightly lower accuracy than PHIST and CHERRY, it offered the advantage of a 100\% prediction rate. In contrast, iPHoP's performance was highly variable; it performed well in the human gut dataset but was less accurate in the wastewater and cow fecal environments, highlighting that environment-specific factors can significantly influence prediction accuracy.

For comparison, we also benchmarked two direct-evidence methods: prophage detection and standard CRISPR spacer matching. As expected, these methods achieved near-perfect accuracy (100\% at the species level) when a prediction could be made, confirming that a direct match is reliable evidence of a virus-host link. This exceptional accuracy, however, was coupled with extremely low prediction rate. The prediction rate for both methods was less than 5\% across the datasets, with CRISPR spacer matching failing to yield any predictions in the wastewater and cow fecal matter samples (Fig. \ref{fig:hic}). This further confirmed our previous finding that significant matches between the viral contigs and host MAGs were rare.

\begin{table*}[h!]
\centering
\caption{The average elapsed time to predict labels of 1,000 viral sequences for each method. All the methods are run on Intel\textsuperscript{\textregistered} Xeon\textsuperscript{\textregistered} Gold 6258R CPU with 40 cores and Tesla A100 (if GPU is required).}
\resizebox{\textwidth}{!}{%
\begin{tabular}{ccccccccccc} \hline
Program & WIsH & CHERRY & iPHoP & PHP & PHERI & vHULK & VHM-Net & DeepHost & PB-LKS & PHIST\\ \hline
Min/1,000 virus & 3 &  5   & 11,981  & 4 & 2,116 & 29 & 240 &  \textless 1 & 10,280 & \textless 1\\ \hline
\end{tabular}%
}
\vspace{-0.3cm}
\label{tab:time}
\end{table*}

\subsubsection{Ensemble of top-performing tools enhances host prediction accuracy} To investigate strategies for maximizing host prediction performance, we ensembled the outputs from the top-performing tools using three distinct strategies (see Methods). The first, ``Union + Majority Vote'', aggregated all predictions and used a majority rule to assign the host. The second, ``Joint + Majority Vote'', also used a majority rule but was restricted to the subset of viruses predicted by all tools. The final and most stringent method, ``Joint + Consensus'', required a unanimous prediction from all three tools to make a final assignment. 

As shown in Supplementary Fig. S6, the ``joint $+$ consensus'' approach leads to a notable improvement, reaching an accuracy of 99\% on the RefSeq-VHDB dataset. The other combination strategies offered no significant benefit over single-tool performance. Consistent with our findings, The "joint + consensus" approach again yielded the best performance, improving average accuracy by up to 15\% over the best individual tool across all three datasets in MetaHiC-VHDB (Supplementary Fig. S7). However, this came at the cost of a significantly lower prediction rate on all datasets. Thus, for specialized applications where accuracy is paramount and lower prediction rate is acceptable, employing a ``joint $+$ consensus'' of top-performers is the optimal strategy to boost the host prediction performance.

\subsubsection{Computational efficiency and scalability of prediction tools}

Beyond prediction performance, the computational efficiency of a tool is a critical factor for its practical application, particularly when analyzing large-scale metagenomic datasets. We, therefore, evaluated the computational performance of each tool by measuring the average elapsed time required to process 1,000 viral contigs on a consistent hardware platform: Intel Xeon Gold 6258R CPU with 40 cores and Tesla A100 if GPU is required.

The results, presented in Table \ref{tab:time}, reveal a stark divergence in computational requirements among the tools. A group of methods, including PHIST, PHP, DeepHost, and CHERRY, demonstrated high efficiency, requiring only a few minutes to complete the task. In contrast, several tools were exceptionally resource-intensive. iPHoP and PB-LKS were the most computationally demanding tools. This significant variation in performance can largely be attributed to the underlying feature extraction process, which is often the primary computational bottleneck. For example, tools like iPHoP rely on searching against an extremely large internal database (over 300 GB in size) to generate features for each prediction. This exhaustive search strategy, while potentially powerful for finding homologous signals, results in a prohibitively long runtime for large-scale analyses. Conversely, the highly efficient tools likely employ more lightweight feature extraction methods, such as $k$-mer frequency calculations or the use of pre-trained models that do not require extensive database searches for every prediction. This trade-off between computational cost and methodological approach may be a key consideration for researchers when selecting a tool appropriate for the scale of their data and available resources.

\section{Discussion}
The rapid expansion of viral sequence data has created a critical need for robust computational methods to link viruses to their microbial hosts. This review provides a comprehensive guide for navigating the complex field of computational virus-host prediction. We began by systematically defining the two primary problem formulations—link prediction and multi-class classification—that frame the task. We then analyzed the genomic features used for prediction, detailing their respective advantages and limitations. Our survey of published tools revealed that while numerous methods exist, their practical accessibility varies significantly. This observation motivated the construction of two distinct benchmark datasets tailored to different research scenarios. The RefSeq-VHDB benchmark was designed to test the classification scenario, where a tool must predict a host for a viral sequence without a specific set of candidate host genomes. In contrast, the MetaHiC-VHDB benchmark simulates a metagenomic analysis, providing both viral contigs and candidate host MAGs from the same environment to directly test a tool's ability to perform de novo link prediction.

A key finding from our benchmarking is that the optimal strategy for host prediction is highly dependent on the research context and the user's specific goals. When selecting a single tool, our results offer a clear guide. On the RefSeq-VHDB dataset, RaFAH demonstrates superior accuracy at higher taxonomic ranks. For discovering novel interactions within metagenomes (MetaHiC-VHDB), link-prediction tools like PHIST and iPHoP are highly effective. CHERRY stands out as a robust generalist, delivering consistent performance across both benchmark types. Beyond the performance of individual tools, our analysis reveals a powerful strategy for applications where precision is paramount. By employing a ``joint $+$ consensus'' ensemble of the top-performing tools, users can increase predictive accuracy to nearly 99\% on reference data. However, this high confidence comes with a significant and predictable trade-off: a sharp reduction in the overall prediction rate. This highlights that users must choose not only the right tool, but also the right approach: For exploratory studies requiring high prediction rate, a high-performing single tool is optimal. For validation or high-confidence discovery where false positives are costly, the joint consensus approach is superior.

Beyond predictive accuracy and prediction rate, our practical evaluation revealed significant barriers that impact a tool's real-world utility. In particular, computational cost is a critical and often-overlooked factor. The substantial resource requirements of tools like iPHoP and PB-LKS can render them impractical for analyzing large-scale metagenomic datasets, whereas the efficiency of methods such as PHIST and PHP makes them highly scalable. Ultimately, such practical considerations are as important as predictive accuracy in determining a tool's widespread adoption by the research community.

Due to the complexity and ongoing evolution of the host prediction problem, this study has limitations inherent to the current state of the field. Primarily, our evaluation framework is constrained by public databases that document one-to-one virus-host associations, and thus assesses the prediction of only a single host per virus. This approach does not fully consider the host range of polyvalent phages and may underestimate the performance of features (e.g., CRISPR spacer match) that correctly identify alternative hosts. However, we found this limitation may have a minimal impact on the validity of our comparative benchmark. The Top-N analysis revealed that considering multiple predictions introduced significant ambiguity for only marginal gains in accuracy. This finding, combined with the understanding that most phages have narrow species-level host ranges, suggests that false negatives from polyvalence are rare and do not skew the overall performance trends. Second, although our analysis covers the most common features in current methods, other signals could inform future approaches, including tRNA gene matching, methylation patterns, and co-abundance profiles. Our preliminary results indicate these features hold potential as complementary signals for host prediction (Supplementary Fig. S8 and S9). Third, while our benchmark provides reliable species-level evaluation, predicting host range at the strain level remains critical for applications like phage therapy. In this context, tools built on a link prediction framework—such as PHP, iPHoP, and CHERRY—offer greater utility due to their ability to output multiple candidates. This need has spurred the development of specialized models for bacteria like Escherichia and Klebsiella, which often incorporate domain-specific knowledge such as O-antigen serotypes to achieve high resolution \cite{boeckaerts2024prediction, gaborieau2024prediction}. This highlights a fundamental trade-off between the broad applicability of general-purpose predictors and the precision of tools tailored to specific microbial systems.

Looking forward, a key challenge is the dynamic nature of virus-host interactions, such as host switching, where a virus's genomic signature may not yet reflect its new tropism. This phenomenon can confound prediction models that depend on established signals of co-living. Fortunately, emerging technologies offer promising avenues to address these dynamics. In particular, single-cell sequencing has demonstrated significant potential to identify rare virus-host interactions and profile phage infections at high resolution \cite{wang2023single}. In contrast to bulk methods like MetaHiC, single-cell approaches can provide more precise data on infection timing, prophage induction, and host gene expression, enabling the study of virus-host dynamics at the level of individual infection events rather than community-level averages.

Finally, while our review focused on viruses that infect prokaryotes, predicting hosts for RNA viruses represents another critical domain with a distinct set of challenges. The compact genomes of RNA viruses offer limited space for host-tropism signals, and the rarity of direct virus-host sequence matches largely precludes the use of link prediction methods. Consequently, existing approaches are typically formulated as multi-class classification tasks that rely on virus-virus similarity \cite{pandit2022predicting, chen2024rnavirhost}. The efficacy of these models, however, is constrained by the high mutation rates of RNA viruses, which reduce the genomic conservation needed for accurate generalization to novel viruses. This confluence of biological hurdles—compounded by the broad host ranges of many RNA viruses and sparse host databases—marks this area as a significant frontier for future research.

\section{Methods}

\subsection{CRISPR spacer, prophage, and tRNA matches}
To establish a labeled data for our analysis, we utilized a dataset of 4,698 viruses with species-level host annotations from the Viral RefSeq database (February 2025). For the host genomic data, we downloaded all 110,988 available prokaryotic genomes from GenBank (February 2025). 

\begin{itemize}
    \item \textbf{CRISPR spacer match}: We obtained 2,005,489 CRISPR spacers derived from the prokaryotic genomes using CRT v1.2 \cite{bland2007crispr}  with default parameters. NCBI BLAST+ v2.16 was employed to align viral genomes to the CRISPR spacers database.
    \vspace{-0.1cm}
    \item \textbf{Prophage match}: we employed NCBI BLAST+ v2.16 to align 4,698 viral genomes to prokaryotic genomes.
    \vspace{-0.1cm}
    \item \textbf{tRNA match}: We obtained 7,709,234 tRNA genes derived from the prokaryotic genomes using ARAGORN \cite{laslett2004aragorn}. Then, NCBI BLAST+ v2.16 was employed to align viral genomes to the tRNA genes database.
    \vspace{-0.1cm}
\end{itemize}

\subsection{Evaluation metrics of CRISPR spacer and prophage matches}

To estimate the usability and reliability of the features derived from CRISPR spacer match and prophage detection, we used four metrics to evaluate the results. We first defined the correct alignment/hit as the viral sequences is aligned to their host or prokaryote has the same taxonomy as its host. Then, the metrics can be calculated as below. To present the statistical process more intuitively, a simple example can be found via Supplementary Fig. S4.

\subsubsection{Top-1 accuracy:} Accuracy summarized on the best hit (the hit with the best alignment score). For each virus we evaluate Top-1 correct: the number of viruses which best hit is the correct hit. Then, the Top-1 correct will be divided by the total number viruses that have alignment results (Eqn. \ref{Eq:top1}).

\vspace{-0.5cm}
\begin{equation}
    \mbox{Top{-}1\ accuracy} = \frac{\mbox{Top{-}1\ correct}}{\mbox{Total\ aligned\ viruses}}
    \label{Eq:top1}
\end{equation}

\subsubsection{Top-N accuracy:} Accuracy summarized on the all the alignment hits. For each virus we evaluate Top-N correct: the number of viruses which correct hit is included in the alignment results. Then, the Top-N correct will be divided by the total number viruses that have alignment results (Eqn. \ref{Eq:topn}).

\vspace{-0.3cm}
\begin{equation}
    \mbox{Top{-}N\ accuracy} = \frac{\mbox{Top{-}N\ correct}}{\mbox{Total\ aligned\ viruses}}
    \label{Eq:topn}
\end{equation}
\vspace{-0.3cm}

\subsubsection{Alignment proportion:} It is a proportion of aligned viruses, which reflect how many viruses can have at least one alignment (Eqn. \ref{Eq:ap}). 

\vspace{-0.3cm}
\begin{equation}
    \mbox{Alignment proportion} = \frac{\mbox{Total aligned viruses}}{\mbox{Total viruses}}
    \label{Eq:ap}
\end{equation}
\vspace{-0.3cm}

\subsubsection{Correct match ratio:} It is a proportion of the correct hits in the alignment results, estimating the errors introduced by using Top-N accuracy. To minimize the bias introduced by the data distribution (some taxonomy has much more prokaryotic genomes), we apply the number of unique taxonomy in the alignment to replace the number of hits. The larger the correct match ratio, the less error introduced by Top-N accuracy (Eqn. \ref{Eq:cmr}).

\vspace{-0.3cm}
\begin{equation}
\begin{split}
    &\mbox{Correct match ratio} = \\
     &\frac{\mbox{Number of virus with a correct hit}}{\sum_i\mbox{Number of unique taxa in virus}_i\mbox{'s alignments}}
     \label{Eq:cmr}
\end{split}
\end{equation}
\vspace{-0.3cm}

\subsection{The RefSeq-VHDB benchmark}
The NCBI RefSeq dataset, initially utilized in VHM and later refined in CHERRY, is a widely adopted benchmark for evaluating virus-host prediction tools. For clarity in this review, we refer to it by the more formal name, the RefSeq Virus-Host Database (RefSeq-VHDB). This dataset is derived from the NCBI Viral RefSeq database and is filtered to retain only viruses with prokaryotic hosts (bacteria and archaea) that have a confident species-level annotation. A key advantage of RefSeq-VHDB is the high confidence of its virus-host linkages and its low data redundancy.

Our version of the dataset, downloaded in February 2025, contains 4,698 viruses linked to 498 distinct host species. The host distribution exhibits a long-tail pattern (Supplementary Fig. S5), a characteristic that poses a significant challenge for model robustness. The three most frequent host species—\textit{Escherichia coli}, \textit{Salmonella enterica}, and \textit{Mycolicibacterium smegmatis}—account for 27.8\% of the entries, while 245 species are represented by only a single virus/interaction.

Previous studies have established two primary methods for evaluating this dataset: a temporal split, which mimics the real-world challenge of classifying newly discovered viruses by training on older entries to predict hosts for newer ones, and a sequence similarity split, which assesses a model's generalization capabilities \cite{ahlgren2017alignment, lu2021prokaryotic, shang2022cherry}. In our benchmark, we chose to use the complete, unsplit dataset. Our rationale was to  realistically reflect the performance users can expect from the latest release versions of all tools.

\subsection{The MetaHiC-VHDB Benchmark}
Evaluating prediction performance in a metagenomic sequencing data is challenging due to the difficulty of establishing ground-truth virus-host pairs. While databases like Prophage-DB exist, they are skewed towards temperate phages. Furthermore, their use for benchmarking can be unfair, as tools incorporating prophage detection steps (e.g., iPHoP, CHERRY, PhageTB) may achieve near-perfect accuracy.

To create a more representative benchmark, we constructed datasets based on proximity ligation sequencing (Hi-C). This method captures physical DNA-DNA interactions, including those between phages replicating within their host cells, providing a robust source of evidence for virus-host linkages. We processed three independent Hi-C sequencing datasets from diverse environments, including human gut \cite{press2017hi}, cow fecal matter \cite{stewart2018assembly}, and wastewater \cite{stalder2019linking} . First, we follow a standard pipeline to process the Hi–C reads as introduced in \cite{wu2023hi}: First, a standard cleaning procedure was applied to all raw WGS and Hi-C read libraries using bbduk from the BBTools suite (v37.62) We discarded short reads below 50 bp at each cleaning step. Adapter sequences were removed by bbduk with parameter ``ktrim=r k=23 mink=11 hdist=1 minlen=50 tpe tbo'' and reads were quality-trimmed using bbduk with parameters “trimq=10 qtrim=r ftm=5 minlen=50.” Then, the first 10 nucleotides of each read were trimmed by bbduk with parameter “ftl=10.”. Second, all identical PCR optical and tile-edge duplicates for Hi-C paired-end reads were removed by the script “clumpify.sh” from BBTools suite. Thrid, the hicstuff (v3.2.4) is employed to generate the contacts matrix on the provided contigs and MAGs from the original research. Then, the resulting virus-host links were subjected to a stringent five-step filtering process to minimize false positives:

\begin{itemize}
    \item Viral contigs were identified using the intersection of predictions from PhaBOX/PhaMer, geNomad, and VirSorter2.
    \vspace{-0.1cm}
    \item A minimum of two Hi-C read pairs were required to link a viral contig to a host MAG.
    \vspace{-0.1cm}
    \item Host MAGs were required to have an intra-MAG connectivity of at least 10 links to ensure they were sufficiently well-assembled.
    \vspace{-0.1cm}
    \item The virus-host connectivity ratio ($R$) had to exceed 0.1.
    \vspace{-0.1cm}
    \item For each virus, only the host with the highest number of supporting Hi-C links was retained as the definitive interaction.
\end{itemize} 

The calculation of connectivity ratio ($R$) is listed in Eqn. \ref{Eq:r}. It is calculated using the connectivity density ($D_{V_iM_j}$) of the virus-MAG pair $(i, j)$ and MAG $j$ to itself ($D_M$), with connectivity density defined as the Hi-C link count per $kb^2$ of sequence. This value is further normalized using a term derived from the viral copy number per MAG $VPM$. The $VPM_{ij}$ is estimated as shown in Eqn. \ref{Eq:vph}, where $v_i$ is the abundance of virus $i$, $M_j$ is the abundance of MAG $j$, $L_{ij}$ is the number of Hi-C links for the specific virus-MAG pair $(i, j)$, and $\sum L(v_i)$ is the sum of links for the virus $i$ with all possible MAGs.

\vspace{-0.3cm}
\begin{equation}
    R_{ij} = \frac{D_{V_iM_j}}{D_{M_j} \times VPH_{ij}}
    \label{Eq:r}
\end{equation}
\vspace{-0.3cm}
\vspace{-0.3cm}
\begin{equation}
    VPH_{ij} = \frac{V_i}{M_j } \frac{L_{ij}}{\sum L(v_i)}
    \label{Eq:vph}
\end{equation}

The taxonomy of all host MAGs was assigned using the GTDB-Toolkit (v2.4.0; Release:R220; classify-wf). The final, curated virus-host pairs from these three environments constitute our MetaHiC benchmark.

\subsection{Evaluation metrics used in benchmarking}
To ensure a fair and comprehensive comparison between existing tools, we selected two widely applicable metrics that can evaluate performance for tools based on both link-prediction and multi-class classification frameworks: Accuracy and Prediction Rate. Also, these two metrics can reflect the performance in real usage scenarios.

\begin{itemize}
\item  \textbf{Accuracy:} Assessing whether the predicted host taxon matches the ground-truth host taxon at a specific taxonomic rank (e.g., species, genus, and family). For the MetaHiC-VHDB benchmark, where specific host MAGs are provided, accuracy can be evaluated at the strain level by determining if the correct MAG was identified.

\item  \textbf{Prediction Rate:} This metric is defined as the percentage of input viruses for which a tool successfully returns a host prediction. Some tools incorporate internal confidence thresholds and may not return a prediction for every query virus if the evidence is deemed insufficient to maintain a high accuracy. This metric, therefore, quantifies the practical applicability and coverage of a tool.
\end{itemize}

To assess whether combining outputs from top-performing tools could improve predictive performance, we evaluated three commonly used ensemble strategies:

\begin{itemize}
\item  \textbf{Union + Majority Vote:} This method aggregates all predictions from the three selected tools. For any virus with multiple predictions, the final host is determined by a majority vote. Ties are resolved by selecting the prediction from the tool with the highest individual performance on our benchmark.

\item  \textbf{Joint + Majority Vote:} This method considers only the subset of viruses for which all three tools provided a prediction. Within this subset, a majority vote is applied to determine the final host, with ties resolved as described above.

\item  \textbf{Joint + Consensus:} The most stringent approach, which yields a prediction only when all three tools are in unanimous agreement on the host assignment.
\end{itemize}

\subsubsection{Data availability} 
All the datasets and supplementary files used in this study are publicly available. The scripts and dataset information can be accessed via GitHub (https://github.com/KennthShang/HostPredictionReview) and Zenodo: (https://doi.org/10.5281/zenodo.16975470). The MetaHiC data can be found via the NCBI Sequence Read Archive database (http://www.ncbi.nlm.nih.gov/sra). The human gut dataset is available under accession codes: shotgun library SRR6131123, Hi-C libraries SRR6131122 and SRR6131124. The cow fecal dataset used in this study is under accession codes: shotgun library ERX2333418, Hi-C libraries ERX2548555 and ERX2548556. The wastewater dataset is available under accession codes: shotgun library SRR8239393 and Hi-C library SRR8239392. 

\subsubsection{Conflict of interest statement} 
The authors declare that they have no competing interests.

\subsubsection{Acknowledgements}
The computation was conducted at the HPCC of City University of Hong Kong. 

\subsubsection{Funding}
This work was supported by the Hong Kong Research Grants Council (RGC) General Research Fund (GRF) [11209823] and City University of Hong Kong (the Institute of Digital Medicine,  9667256, 9678241).

\bibliographystyle{unsrt}  
\bibliography{references}

\end{document}